\documentclass[conference,10pt]{IEEEtran}

\usepackage{amsmath, mathrsfs}

\usepackage{stmaryrd}
\usepackage{relsize}

\usepackage{cite}
\usepackage{amsfonts}
\usepackage{amssymb}
\usepackage{color}
\usepackage{multirow}
\usepackage{graphicx}
\usepackage{relsize}
\usepackage{multirow}
\usepackage{multicol}
\usepackage{caption}
\usepackage{subcaption}
\usepackage[numbers,sort&compress]{natbib}
\usepackage{balance}
\usepackage{arydshln}
\allowdisplaybreaks

\usepackage{amsfonts}
\usepackage{amssymb}
\usepackage{graphicx}
\usepackage{color}
\usepackage{multirow}
\usepackage{graphicx}

\usepackage{caption}
\usepackage{subcaption}
\usepackage[numbers,sort&compress]{natbib}

\usepackage{algorithm}
\usepackage[noend]{algpseudocode}

\usepackage{lipsum}

\newcommand\blfootnote[1]{%
	\begingroup
	\renewcommand\thefootnote{}\footnote{#1}%
	\addtocounter{footnote}{-1}%
	\endgroup
}

\newcommand{\subparagraph}{}
\usepackage[compact]{titlesec}
\titlespacing*{\section}{15pt}{1\baselineskip}{0.9\baselineskip}

\allowdisplaybreaks

\newcommand{\myhash}{%
  {\settoheight{\dimen0}{C}\kern-.05em\, \resizebox{!}{\dimen0}{\raisebox{\depth}{\#}}}}

\usepackage{multicol}


\usepackage{mathbbol}

\def\mindex#1{\index{#1}}

\def\sq{\hbox{\rlap{$\sqcap$}$\sqcup$}}
\def\qed{\ifmmode\sq\else{\unskip\nobreak\hfil
\penalty50\hskip1em\null\nobreak\hfil\sq
\parfillskip=0pt\finalhyphendemerits=0\endgraf}\fi\medskip}

\long\def\defbox#1{\framebox[.9\hsize][c]{\parbox{.85\hsize}{%
\parindent=0pt
\baselineskip=12pt plus .1pt      
\parskip=6pt plus 1.5pt minus 1pt 
 #1}}}

\long\def\beginbox#1\endbox{\subsection*{}%
\hbox{\hspace{.05\hsize}\defbox{\medskip#1\bigskip}}%
\subsection*{}}

\def\endbox{}

\newsavebox{\junk}
\savebox{\junk}[1.6mm]{\hbox{$|\!|\!|$}}

\def\Re{\field{R}}

\def\bC{{\mathbb C}}

\def\bR{{\mathbb R}}

\def\bfmath#1{{\mathchoice{\mbox{\boldmath$#1$}}%
{\mbox{\boldmath$#1$}}%
{\mbox{\boldmath$\scriptstyle#1$}}%
{\mbox{\boldmath$\scriptscriptstyle#1$}}}}

\def\bfmY{\bfmath{Y}}

\def\bfmhhaY{\bfmath{\hhaY}} 
\def\bfmhhaY{\hbox to 0pt{$\widehat{\bfmY}$\hss}\widehat{\phantom{\raise 1.25pt\hbox{$\bfmY$}}}}

\def\til={{\widetilde =}}

 \def\FRAC#1#2#3{\genfrac{}{}{}{#1}{#2}{#3}}

\def\ddtp{{\mathchoice{\FRAC{1}{d^{\hbox to 2pt{\rm\tiny +\hss}}}{dt}}%
{\FRAC{1}{d^{\hbox to 2pt{\rm\tiny +\hss}}}{dt}}%
{\FRAC{3}{d^{\hbox to 2pt{\rm\tiny +\hss}}}{dt}}%
{\FRAC{3}{d^{\hbox to 2pt{\rm\tiny +\hss}}}{dt}}}}

\def\average#1,#2,{{1\over #2} \sum_{#1}^{#2}}

\def\eye(#1){{\bf(#1)}\quad}

\def\eq#1/{(\ref{e:#1})}

\newcommand{\beqn}[1]{\notes{#1}%
\begin{eqnarray} \elabel{#1}}

\newcommand{\eeqn}{\end{eqnarray} }

\newcommand{\beq}[1]{\notes{#1}%
\begin{equation}\elabel{#1}}

\newcommand{\eeq}{\end{equation}}

\def\bdes{\begin{description}}
\def\edes{\end{description}}

\newcounter{rmnum}

\newcounter{anum}

{\end{list}}

\def\ass(#1:#2){(#1\ref{#1:#2})}

\def\ritem#1{
\item[{\sf \ass(\current_model:#1)}]
}

\newenvironment{recall-ass}[1]{%
\begin{description}
\def\current_model{#1}}{
\end{description}
}

\usepackage{tikz}
\usepackage{pgfplots,tikz-3dplot}
\pgfplotsset{compat=newest}
\usetikzlibrary{patterns}
\usetikzlibrary{positioning}
\usetikzlibrary{datavisualization}
\usetikzlibrary{datavisualization.formats.functions}
\usetikzlibrary{backgrounds}
\usetikzlibrary{shapes,snakes}

\usepgfplotslibrary{external} 
\usepackage{float}
\usepackage{afterpage}
\usepackage{balance}

\usepackage{geometry}
\newtheorem{problem}{{\it Problem}}

\long\def\comment#1{}

\newfont{\bb}{msbm10 scaled 1100}

\newcommand{\av}{{\bf a}}
\newcommand{\bv}{{\bf b}}
\newcommand{\cv}{{\bf c}}

\newcommand{\hv}{{\bf h}}

\newcommand{\pv}{{\bf p}}

\newcommand{\rv}{{\bf r}}

\newcommand{\uv}{{\bf u}}
\newcommand{\wv}{{\bf w}}

\newcommand{\xv}{{\bf x}}
\newcommand{\yv}{{\bf y}}
\newcommand{\zv}{{\bf z}}

\newcommand{\Am}{{\bf A}}

\newcommand{\Cm}{{\bf C}}
\newcommand{\Dm}{{\bf D}}

\newcommand{\Dc}{{\cal D}}

\newcommand{\Fc}{{\cal F}}
\newcommand{\Gc}{{\cal G}}

\newcommand{\Ic}{{\cal I}}

\newcommand{\Pc}{{\cal P}}

\newcommand{\Tc}{{\cal T}}

\newcommand{\Xc}{{\cal X}}

\newcommand{\Phim}{\boldsymbol{\Phi}}

\renewcommand{\arg}{{\hbox{arg}}}

\renewcommand{\Re}{{\rm Re}}
\renewcommand{\Im}{{\rm Im}}

\newcommand{\transp}{{\sf T}}

\newcommand{\taum}{\tau_{\text{max}}}
\newcommand{\tauep}{\tau_\epsilon}

\graphicspath{ {./Figures/} }

\begin{document}
\title{WiFi-Based Indoor Localization via Multi-Band Splicing and Phase Retrieval}
\author{\IEEEauthorblockN{Mahdi Barzegar Khalilsarai\IEEEauthorrefmark{1},
Stelios Stefanatos\IEEEauthorrefmark{2}, Gerhard Wunder\IEEEauthorrefmark{2}, and Giuseppe Caire\IEEEauthorrefmark{1}}
\vspace{-4mm}\\
$^{\ast}$Communications and Information Theory Group, Technische Universit\"{a}t Berlin\\
$^{\dagger}$Heisenberg Communications and Information Theory Group, Freie Universit\"{a}t Berlin\\
Emails: $\{$m.barzegarkhalilsarai, caire$\}$@tu-berlin.de, $\{$stelios.stefanatos, g.wunder$\}$@fu-berlin.de}
\maketitle
\begin{abstract}
We study the problem of indoor localization using commodity WiFi channel state information (CSI) measurements. The accuracy of methods developed to address this problem is limited by the overall bandwidth used by the WiFi device as well as various types of signal distortions imposed by the underlying hardware. In this paper, we propose a localization method that performs channel impulse response (CIR) estimation by splicing measured CSI over multiple WiFi bands. In order to overcome hardware-induced phase distortions, we propose a phase retrieval (PR) scheme that only uses CSI magnitude values to estimate the CIR. To achieve high localization accuracy, the PR scheme involves a sparse recovery step, which exploits the fact that the CIR is sparse over the delay domain, due to the small number of contributing signal paths in an indoor environment. Simulation results indicate that our approach outperforms the state of the art by an order of magnitude (cm-level localization accuracy) for more than 90\% of the trials and for various SNR regimes.
\end{abstract}
\vspace{-1mm}
\section{Introduction}
Multipath channel estimation using WiFi frequency domain channel state information (CSI) has found numerous applications in indoor localization and ranging in recent years \cite{kotaru2015spotfi,vasisht2016decimeter,xie2018precise,zhuo2017perceiving}. Such applications rely on the channel delay profile to estimate the distance between two devices or detect the motion of an object. The accuracy of such schemes is limited by the resolution with which they can estimate the delay components of the channel. This resolution is limited by the inverse of the overall bandwidth of the WiFi device. For example, using CSI measurements over a bandwidth of $W=20$ MHz, one can localize channel delay components separated by $\Delta \tau = \frac{1}{W}=50$ ns, whereas when the bandwidth increases to $1$ GHz, one can improve the resolution to $\Delta \tau = 1$ ns \cite{xie2018precise}. Multiplying the delay resolution with the speed of the electromagnetic wave, results in a ranging error of up to $15$ m in the former versus up to only $30$ cm in the latter case. Therefore, a large channel bandwidth is crucial for any application demanding precise localization. \blfootnote{This work was carried out within the ``Secure Fog-Connection Layer for IoT Applications (SecureFog)" project,  funded by the Federal Ministry of Education and Research (BMBF), Germany, under grant number FKZ: 16KIS0776.}

Since single WiFi bands usually span over $20$ MHz, using each of them separately results in poor localization. However, the total bandwidth allocated to a WiFi device, which consists of all the used WiFi bands, may scale up to several hundred MHz or even multi-GHz, making WiFi CSI measurements over the wide bandwidth a useful resource for decimeter-level localization tasks. The process of exploiting such measurements over multiple bands is referred to as multi-band splicing \cite{xie2018precise}. Despite the potential of multi-band splicing, channel impulse response (CIR) estimation using raw WiFi CSI measurements is challenging, mainly due to imperfections of the underlying hardware. Per-band CSI measurements are performed independently based on sequential frequency hopping, which introduces random and independent phase offsets to the measurements at each band. Furthermore, time synchronization errors introduce a linear phase rotation to the CSI, which is independent from one band to the other \cite{vasisht2016decimeter}. 

The existence of such distortions makes a straightforward CIR estimation via multi-band splicing impossible. A few methods have been recently proposed to solve this problem, by removing the phase distortions using signal processing tools. In \cite{vasisht2016decimeter}, the Chronos system was proposed, which uses the CSI measured only on a single, specially chosen subcarrier per band in order to recover the time of flight (ToF), which is the delay corresponding to the first path, assuming a sparse multipath channel. Using a single subcarrier per band may lead to a poor estimation of the ToF, since for channels with a large delay spread, it results in an insufficient sampling rate and an aliased estimation of the CIR. In addition, the CSI measurements over the remaining subcarriers of a band can provide valuable information, especially in low signal-to-noise ratio (SNR) scenarios, but are neglected. The problem of spectrum splicing with the application of accurate ToF estimation was also treated in \cite{xie2018precise} and \cite{zhuo2017perceiving}, where heuristic methods were proposed with overall performances inferior to the method of \cite{vasisht2016decimeter}, which we take as the state of the art for comparison to our work.

In this paper we develop a method for multi-band splicing with phase-distorted and noisy WiFi CSI measurements. This method operates only on the magnitude of the channel frequency samples over multiple WiFi bands to estimate the CIR, a technique which is known as phase retrieval (PR) in the literature \cite{candes2013phaselift}. As is well known, applying PR to frequency samples results in a fundamental ambiguity in the estimated delay domain signal, in terms of constant shifts and conjugate-reflections with respect to the origin \cite{jaganathan2017sparse}. We resolve this ambiguity using a \textit{handshaking} process between the transmitter and the receiver and exploiting the channel reciprocity \cite{guillaud2005practical}. We show that our method can recover the channel delay components and their coefficients with a high resolution for various SNR values, which results in significant ranging accuracy. 

Notation: for a positive integer $M$ we denote by $[M]$ the set \scalebox{0.9}{$\{ 0,\ldots,M-1 \}$}. For a vector $\xv$, $|\xv|$ denotes elementwise absolute values. For a function $h(\tau)$, we define the Fourier transform as \scalebox{0.9}{$\Fc \{ h(\tau) \}|_f = \int h(\tau) e^{-j2\pi \tau f} d\tau$}. For a set of integers $J$ and a vector $\xv$, $\xv_J$ denotes the elements of $\xv$ in $J$.
\section{System Setup}\label{sec:sys_set}
Assume that the propagation environment between a WiFi transmitter and a receiver consists of $K$ scatterers, each imposing a path gain $c_k\in \bC$ and delay $\tau_k \in \Xi \overset{\Delta}{=}[0,\taum]$, $k\in [K]$ where $\taum>0$ denotes the maximum delay spread. Using OFDM signaling, the WiFi device transmits pilots to the receiver over $M$ frequency bands, each consisting of $N$ equi-spaced subcarriers.\footnote{Note that the bands need not be adjacent, as is sometimes the case where the WiFi device uses multiple bands around $2$ GHz and $5$ GHz \cite{vasisht2016decimeter}.} The receiver collects the CSI of each band separately, by down-conversion through removing the carrier frequency. The baseband representation of the CIR corresponding to band $m\in [M]$ is thereby given as $h^{(m)} (\tau) = \sum_{k=0}^{K-1} c_k e^{-j2\pi f_{m,0} \tau_k}\delta (\tau-\tau_k),$ where $f_{m,0}$ denotes the carrier frequency of the $m$-th band and $\delta (\cdot)$ denotes Dirac's delta. Since only a few propagation paths contribute to the channel, i.e. $K$ is a small number, the CIR admits a sparse representation over the delay domain, which will be exploited in our proposed method. The channel frequency response (CFR) samples over the set of subcarriers $\{ n\, f_s : n= -\tfrac{N-1}{2},\ldots,\tfrac{N-1}{2}\}$\footnote{$N$ is an odd integer.}, with $f_s$ being the carrier spacing, are denoted by the vector $\tilde{\hv}^{(m)}\in \bC^{N}$ where 
\begin{equation}\label{eq:freq_samples}
\tilde{\hv}^{(m)}_n= \sum_{k=0}^{K-1} c_k e^{-j2\pi (f_{m,0}+nf_s)\tau_k }= \sum_{k=0}^{K-1} c_k e^{-j2\pi f_{m,n}\tau_k },
\end{equation}
\vspace{-2mm}

 \noindent where $f_{m,n}\overset{\Delta}{=}f_{m,0}+nf_s$. 
 
 Unfortunately, these (raw) CSI measurements are subject to several phase and magnitude distortions as well as additive noise due to hardware imperfections. We denote the CSI of band $m\in [M]$ by the vector $\yv^{(m)}\in \bC^{N}$ where each element can be expressed as \cite{vasisht2016decimeter,xie2018precise,speth1999optimum}
\begin{equation}\label{eq:raw_CSI}
\yv_n^{(m)} = \alpha e^{-j \phi_{m,n} } \tilde{\hv}^{(m)}_n+ \zv_n^{(m)},
\end{equation}
\vspace{-5mm}

\noindent where $\phi_{m,n}\overset{\Delta}{=} 2\pi n f_s \delta_m  + \psi_m, $ is an affine phase distortion term with $\delta_m$ denoting the band-specific time offset due to the packet detection delay (PDD) and $\psi_m$ denoting the constant phase offset. Also $\zv^{(m)} \in \bC^{N}$, by assuming that the support\footnote{The support of a function $f(x)$ denotes the set $\{ x : f(x)\neq 0 \}$.} of the effective CIR (including the shift caused by the PDD) is contained within the OFDM cyclic prefix, is the additive white Gaussian noise (AWGN) vector, and $\alpha>0$ is the power control gain. For simplicity, we set $\alpha=1$ in what follows, since its effect can be subsumed to the noise level.
We refer the interested reader to references \cite{vasisht2016decimeter,xie2018precise,zhuo2017perceiving} for an extensive account on the physical source of these distortions and focus on the mathematical presentation of the problem. 
\subsection{Denoising and Zero-Subcarrier CSI Estimation}\label{sec:interp}
Before proceeding with the statement of the problem, we should address an issue with the raw WiFi measurements. The WiFi devices do not transmit on the zero-subcarrier of each band, i.e. on the frequencies $f_{m,0},~m\in [M]$, since measurements on these subcarriers overlap with the hardware DC offsets that are hard to remove \cite{vasisht2016decimeter}. This means that in each band $m$, the measurement $\yv^{(m)}_0$ is not observed. However, the CSI on the zero subcarrier is very useful, since the phase distortion term $\phi_{m,0},~m\in [M]$ is free of PDD and reduces to a constant offset, i.e., $\phi_{m,0} = \psi_m$. This property will be exploited in the final step of our proposed method. Here, we use the measured CSI over the remaining subcarriers of the band, to interpolate the missing CSI measurement on the zero-subcarrier. 

The authors of \cite{vasisht2016decimeter} used a cubic spline interpolator for this task. Motivated by the fact that the CIR is sparse over the delay domain ($K\ll N$), we propose a different interpolation/estimation scheme. Our estimator is based on $\ell_1$-norm minimization, which exploits channel sparsity and will additionally denoise the raw CSI measurements. Recalling \eqref{eq:raw_CSI}, one can write
$\yv^{(m)} = \sum_{k=0}^{K-1} w_k^{(m)}  \bv^{(m)} (\tau_k) +\zv^{(m)},$ where $\bv^{(m)}_n (\tau_k) = \exp\left(-j2\pi \left( f_{m,n}\tau_k+n f_s\delta_m \right) \right)$ and $w_k^{(m)}=c_k e^{-j\psi_m}$.
Since the signal component of $\yv^{(m)}$ is a linear combination of a small number ($K$) of the vectors $\bv^{(m)} (\tau_k)$, it is natural to use a sparse recovery method for both estimating $\yv_0^{(m)}$ and denoising the observed values of $\yv^{(m)}$. Here we use the basis pursuit denoising (BPDN) method for this purpose. Define the oversampled DFT matrix $\Dm\in \bC^{N\times L}$ where $[\Dm]_{n,\ell} = \frac{1}{\sqrt{N}} e^{-j 2 \pi \frac{n\ell}{L}} ~,n\in[N],\,\ell \in [L]$ and $L\gg N$. This matrix serves as a dictionay over a discretized delay grid with $L$ delays over $[0,1]$. The BPDN program, thereby, can be formulated as follows \cite{eldar2012compressed}
\begin{equation}\label{eq:l1_denoising_band}
\begin{aligned}
\tilde{\wv}^{(m)} =  \underset{\wv \in \bC^{L}}{\arg \min}~
\frac{1}{2}\Vert \yv_J^{(m)} - \left( \Dm \wv \right)_J \Vert_2^2  + \rho \Vert \wv \Vert_1, \\
\end{aligned}
\end{equation}
where $\Vert \cdot \Vert_1$ denotes the $\ell_1$ norm, $\rho>0$ is an appropriate regularization scalar and $J = \{-\tfrac{N-1}{2},\ldots,\tfrac{N-1}{2}\}\backslash \{0\}$. In practice $J$ can be any subcarrier index set for which the measurements are available (i.e., pilot symbols are effectively transmitted). Solving \eqref{eq:l1_denoising_band} gives us an estimate of the sparse coefficients vector $\tilde{\wv}^{(m)}$, using which we can not only estimate the zero subcarrier CSI as $\left( \Dm \tilde{\wv}^{(m)} \right)_0$, but also denoise the observed vector $\yv^{(m)}_J$. We denote the overall denoised CSI vector over band $m\in [M]$ as $\yv^{(m)}_{\sf den}=\Dm \tilde{\wv}^{(m)}$. 
The above program is convex and is solved for each band separately, using any of the numerous convex solvers. For a fast implementation we use the method proposed in \cite{wright2009sparse}, known as SpaRSA. Now, by having (denoised) measurements over all subcarriers in all bands, we are ready to state the problem of CIR estimation.\\
\vspace{-4mm}
\subsection{Problem Statement}
Let $\tilde{\hv} \overset{\Delta}{=} [ (\tilde{\hv}^{(0)})^\transp,\ldots,(\tilde{\hv}^{(M-1)})^\transp ]^\transp\in \bC^{MN}$ denote the vector containing the CFR samples over all bands. Using \eqref{eq:freq_samples} one can easily see that $\tilde{\hv}$ contains the frequency samples of the CIR $h(\tau) = \sum_{k=0}^{K-1}c_k \delta (\tau - \tau_k)$ over the ordered set of subcarrier frequencies $ \{f_{0,-\tfrac{N-1}{2}},\ldots,f_{M-1,\tfrac{N-1}{2}}\}$. In the same way we define $\yv  \overset{\Delta}{=}[ (\tilde{\yv}_{\sf den}^{(0)})^\transp,\ldots,(\tilde{\yv}_{\sf den}^{(M-1)})^\transp ]^\transp\in \bC^{MN}$ and $\zv \overset{\Delta}{=} [ (\tilde{\zv}^{(0)})^\transp,\ldots,(\tilde{\zv}^{(M-1)})^\transp ]^\transp\in \bC^{MN} $ as the denoised measurements vector and the noise vector, respectively. It follows that the CSI measurement vector can be written as $\yv =  \Phim\tilde{\hv} + \zv,$ where, recalling \eqref{eq:raw_CSI}, $\Phim \in \bC^{MN\times MN}$ is an \textit{unknown} diagonal matrix, with unit modulus phase-distortion entries. The main problem addressed in this paper is formally stated as follows.
\begin{problem}\label{problem:CSI_estimation}
	Given the distorted CSI measurements vector $\yv$, estimate the sparse CIR $h(\tau)=\sum_{k=0}^{K-1} c_k \delta (\tau - \tau_k)$. 
\end{problem}
Since the samples are only distorted in phase, and not in magnitude, our idea in this paper is to estimate the CIR by applying a PR algorithm to the magnitude of the CSI, i.e. $\{|\yv_i|^2\}_{i\in [MN]}$. An important feature of formulating and solving the problem as such is that, unlike previous methods in the literature, we do not perform any time-consuming pre-processing to remove the effect of phase distortions and we are able, in principle, to estimate the CIR as soon as a single snapshot of CSI measurement across all bands is collected. This is crucial in delay-sensitive applications. 
\section{CIR Estimation via Phase Retrieval}\label{sec:PR_CIR_Estimation}
Using the formulation developed in the previous section, define
\vspace{-5mm}

\begin{equation}\label{eq:u_formula}
\uv_i \overset{\Delta}{=}| \yv_i |^2 = | \tilde{\hv}_i |^2 + \tilde{\zv}_i,
\end{equation}
\vspace{-5mm}

\noindent where $\uv$ denotes the vector of magnitudes and $\tilde{\zv}_i\overset{\Delta}{=}\tilde{\hv}_i \Phim_{i,i}^\ast \zv_i + \tilde{\hv}_i^\ast\Phim_{i,i} \zv_i^\ast +  |\zv_i|^2$. It is easy to see that $| \tilde{\hv}_i |^2, i\in [MN] $ represents the samples of the Fourier transform of the CIR autocorrelation function
\begin{equation}\label{eq:g_func}
R(\xi) \overset{\Delta}{=} h(\tau) \star h^\ast (-\tau)|_\xi= \sum_{k=0}^{K-1} \sum_{\ell =0 }^{K-1} c_k c_\ell^\ast \delta ( \xi - (\tau_k-\tau_\ell) ),
\end{equation}
\vspace{-3mm}

\noindent for $\xi\in [-\taum,\taum]$, where $\star$ denotes convolution. In order to recover $h(\tau)$ from $\{ \uv_i \}_{i\in [MN]}$ we first recover $R(\xi)$ and then estimate the set of delays $\{ \tau_k \}_{k\in[K]}$ and their corresponding coefficients $\{|c_k|\}_{k\in [K]}$ (up to a phase constant) from the estimated autocorrelation.
\subsection{Sparse Recovery of the CIR Autocorrelation}\label{sec:sp_autocorr_rec}
From \eqref{eq:g_func} it follows that the autocorrelation $R(\xi)$ is conjugate symmetric with respect to $\xi=0$. We can write
$ R(\xi) = r_0 \delta(\xi) + \sum_{s=1}^{\tilde{K}} r_s \delta (\tau - \xi_s ) +  \sum_{s=1}^{\tilde{K}} r_s^\ast \delta (\tau + \xi_s )$ where $\tilde{K}\overset{\Delta}{=}\frac{K(K-1)}{2}$, and $\xi_s \overset{\Delta}{=} |\tau_k - \tau_\ell| \in \Xi $ for some $k,\ell \in [K]$. Furthermore, $R(\xi)$ is a sparse function, i.e. assuming $K$ to be small, it consists of a few number of delta functions over $\Xi \cup \{ -\Xi \}$. Using \eqref{eq:u_formula} we can write
\begin{equation}
\begin{aligned}
\scalebox{0.94}{$\uv$} &\scalebox{0.94}{$= r_0 \mathbf{1} + \sum_{s=1}^{\tilde{K}} \left( r_s \av (\xi_s) + r_s^\ast \av(-\xi_s )\right) + \tilde{\zv}$}\\
& \scalebox{0.94}{$= r_0 \mathbf{1} + 2 \sum_{s=1}^{\tilde{K}} \left( \Re \left\{ r_s\right\}  \av_{\text{re}} (\xi_s)  - \Im \left\{ r_s\right\}  \av_{\text{im}} (\xi_s)\right) + \tilde{\zv},$}
\end{aligned}
\end{equation}
where $\av (\xi_s)=[e^{-j2\pi f_0 \xi_s},\ldots,e^{-j2\pi f_{MN-1} \xi_s}]^\transp$, $\av_{\text{re}} (\xi_s) = \Re \{ \av (\xi_s) \}$, $\av_{\text{im}} (\xi_s) = \Im \{ \av (\xi_s) \}$ and $\mathbf{1}$ denotes the all ones vector of size $MN$. In order to recover $R(\xi)$, we define a dictionary of the vectors $\av (\xi)$ on a dense grid of points over $\Xi$ and apply sparse recovery methods. Let $\Gc = \{ \tfrac{\taum}{G},\tfrac{2\taum}{G},\ldots, \tfrac{G\taum}{G} \}$ with $G\gg MN$ be a dense grid over $\Xi$. Define the dictionary $\Am \overset{\Delta}{=}\left[ \mathbf{1},\,2\Am_1\, ,-2\Am_2  \right]$ where $\Am_1 = \Re \left\{\left[ \av (\xi_0),\ldots, \av (\xi_{G-1}) \right]\right\}\in \bR^{MN\times G}$ and $\Am_2 = \Im \left\{\left[ \av (\xi_0),\ldots, \av (\xi_{G-1}) \right]\right\}\in \bR^{MN\times G},~\xi_i \in \Gc$. We can approximate the signal part of $\uv$ with a linear combination of the columns of $\Am$ and therefore we have $|\tilde{\hv}|^2 \approx \Am \xv,$ where $\xv = [x_0, \, \xv_{\text{re}}^\transp,\xv_{\text{im}}^\transp ]^\transp$ is a sparse vector, where $\xv_{\text{re}}\in \bR^{G}$ and $\xv_{\text{im}} \in \bR^{G}$ denote the real and imaginary parts of the coefficients approximating the frequency samples vector of the autocorrelation, respectively. In order to estimate the sparse vector $\xv$ given $\uv$, we solve the following BPDN problem,
\vspace{-3mm}

\begin{equation}\label{eq:sp_R_est}
\xv^\star = \underset{\xv \in \bC^{2G+1}}{\text{minimize}}~\frac{1}{2}\Vert \Am \xv - \uv \Vert_2^2  + \lambda \Vert \xv \Vert_1,
\end{equation}
\vspace{-5mm}

\noindent where $\lambda>0$ is an appropriate regularization scalar which controls the balance between sparsity of the solution and the approximation accuracy. This parameter depends on the noise level and can be chosen empirically. As explained in Section \ref{sec:interp}, the BPDN problem is a convex program and can be efficiently solved using any convex solver. Similar to problem \eqref{eq:l1_denoising_band}, we solve \eqref{eq:sp_R_est} using the SpaRSA algorithm.  

Once we solve \eqref{eq:sp_R_est}, we have an estimate of a discrete approximation of $R(\xi)$. Using the notation $\xv^\star = [x_0^\star, \, (\xv_{\text{re}}^\star)^\transp,(\xv_{\text{im}}^\star)^\transp ]^\transp$, the coefficients of this discrete approximation over $(0,\taum]$ are given by $\hat{R}(\xi_i) = [\xv_{\text{re}}^\star]_i + j [\xv_{\text{im}}^\star]_i $ for $\xi_i \in \Gc,$ and $\hat{R}(\xi_0)=x^\star_0$, whereas the coefficients over $[-\taum,0)$ are simply given by $\hat{R}(-\xi_i)=\hat{R}(\xi_i)^\ast$. 

This discretized approximation of the autocorrelation usually contains spurious non-zero elements, since the true support entries $\Dc=\{\xi_s\}_{s=1}^{\tilde{K}}$ do not necessarily lie on the grid $\Gc$ and also due to noise effects. This means that the support set of $\hat{R}(\xi)$, i.e. $\{ \xi : \hat{R}(\xi)\neq 0 \}$ generally has more than $2\tilde{K}+1$ elements. However, by knowing the number of paths $K$ contributing to the CIR,\footnote{The number of paths corresponds to the slow-varying geometry of the environment and can be estimated offline during a pre-processing step.} we can refine the support of $\hat{R}(\xi)$ as follows. Let $\Ic = \{ \xi_i>0 : |\hat{R}(\xi)| \ge \varepsilon \}$ denote the set of points in $\Gc$ whose corresponding coefficients in $R(\xi)$ are large enough, where $\varepsilon>0$ is a small, predefined threshold. The points in $\Ic$ are clusters around the true support points $\{ \xi_s \}_{s=1}^{\tilde{K}}$. Knowing the value $\tilde{K}$, we apply the k-means clustering method \cite{bishop2006pattern}  and obtain the center points of these clusters, denoting them by $\{ \hat{\xi}_s \}_{s=1}^{\tilde{K}}$. Eventually, the estimated autocorrelation support is given by $\hat{\Dc} = \{ \hat{\xi}_0 = 0 \} \cup \{\pm \hat{\xi}_s \}_{s=1}^{\tilde{K}}$. Now, the coefficients corresponding to points in the estimated support set can be obtained by solving a least-squares problem 
\begin{equation}
\rv^\star = \underset{\rv\in \bC^{2\tilde{K}+1}}{\arg \min}~\Vert \Am_{\hat{\Dc}}\rv - \uv \Vert_2
\end{equation}
\vspace{-3mm}

\noindent where $\Am_{\hat{\Dc}} = \left[ \av (-\hat{\xi}_{\tilde{K}}),\ldots,\av (0),\ldots, \av (\hat{\xi}_{\tilde{K}})\right]$. In what follows we refer to the estimated coefficient corresponding to $\hat{\xi}_s,~s=1,\ldots,\tilde{K}$, as $\hat{r}_s$. Note that the coefficient corresponding to $-\hat{\xi}_s$ will be $\hat{r}_s^\ast$ due to conjugate symmetry. This completes the estimation of the autocorrelation function $R(\xi)$. 
\subsection{CIR Recovery}
In order to recover the CIR $h(\tau)$ from its autocorrelation estimate $\hat{R}(\xi)$, we rely on a method developed in \cite{baechler2018super} which performs support and magnitude recovery in succession. First, note that the support of the autocorrelation $\Dc= \{ \tau_k - \tau_\ell \}_{k,\ell \in [K]}$ is the difference set of the support of the CIR $\Tc = \{\tau_k\}_{k\in [K]}$, i.e. $\Dc = \Tc - \Tc$ (Minkowski difference). Also $\Dc$ is a symmetric set containing the point $0$. Now we have an estimate of $\Dc$ and want to recover $\Tc$.
A crucial point here is that, recovering $\Tc$ from $\hat{\Dc}$ involves a fundamental ambiguity even when $\hat{\Dc} = \Dc$: for any constant $b\in \bR$ the generic sets $b+\Tc$ and $b-\Tc$ have the same difference set as the $\Tc$. Thus, we can hope to recover $\Tc$ only up to constant \textit{shifts} and \textit{reflections} with respect to 0. These uncertainties will be resolved in a later step of our algorithm and in this section, we only focus on estimating $\Tc$ up to the mentioned ambiguities. We further assume that there exists no collision in the difference set, i.e. there are no two distinct pair of points $(\tau_k,\tau_\ell)$ and $(\tau_{k'},\tau_{\ell'})$ for which $\tau_k-\tau_\ell = \tau_{k'}-\tau_{\ell'}$, which is almost surely guaranteed given that the support elements are driven uniformly at random over the delay domain. The support recovery algorithm is summarized in Algorithm \ref{Alg:supp_rec} and the interested reader is referred to \cite{baechler2018super} for further details.
\begin{algorithm}[t]
	\caption{CIR Support Estimation}\label{Alg:supp_rec}
	\begin{algorithmic}[1]
		\State \textbf{Input:} Estimated autocorrelation support $\hat{\Dc} = \{ \hat{\xi}_s \}_{s=1}^{\tilde{K}}$ 
		\State Initialize the sets $\Xc_2 = \{ 0 , \hat{\xi}_1 \}$ and $\Pc_2 = \hat{\Dc}\backslash \Xc_2$.
		\For{$k =2$ to $K-1$}                    
		\State Select $\hat{\xi}_i\in \Pc_k$ such that $\{ \Xc_k\cup \hat{\xi}_i \} - \{ \Xc_k\cup \hat{\xi}_i \} \subseteq \hat{\Dc}$
		\State $\Xc_{k+1}=\Xc_k \cup \hat{\xi}_i$
		\State $\Pc_{k+1}=\hat{\Dc}\backslash\Xc_{k+1}$
		\EndFor \textbf{end}
		\State {\small $\Xc_K\leftarrow \Xc_K - \min \{ \Xc_K \}$}  \Comment{Shift the solution set such $~~~~~~~~~~~~~~~~~~~~~~~~~~~~~~~~~~~~~~~$that $\min\{\Xc_K\}=0$.}
		\State \textbf{Output:} $\hat{\Tc} = \Xc_K$
	\end{algorithmic}
\end{algorithm}
Once we obtain the CIR support estimate $\hat{\Tc}$, we proceed with estimating the CIR coefficients $\cv = [c_1,\ldots,c_k]^\transp$. Estimation of the coefficients from the autocorrelation would be up to a common constant phase, since it is easy to show that the CIR coefficients $\cv$ and $e^{j\phi}\cv$, result in the same autocorrelation. For any two elements $\hat{\tau}_k,\hat{\tau}_\ell \in \hat{\Tc}$ we can find their corresponding differences $ \pm \hat{\xi}_s = \pm |\tau_k - \tau_\ell |$ in $\hat{\Dc}$ and furthermore, the corresponding estimated coefficients $\hat{r}_s,~\hat{r}_s^\ast$ from the discussion of Section \ref{sec:sp_autocorr_rec}. From \eqref{eq:g_func}, we know that $|r_s| = |c_k c_\ell^\ast|$ and since $\hat{r}_s$ is an estimate of $r_s$ we expect $|\hat{r}_s|= |\hat{c}_k\hat{c}_\ell^\ast|$.
Now construct the matrix $\Cm \in \bR^{K\times K}$ such that
\begin{equation}
\Cm_{k,\ell} \overset{\Delta}{=} 
\begin{cases}
0 &\quad  k=\ell,\\
\log |\hat{r}_{s}| = \log |\hat{c}_k| + \log |\hat{c}_\ell| & \quad k\neq \ell.
\end{cases}
\end{equation}
Note that $\sum_{\ell=1}^{K} \Cm_{k,\ell} = (K-2) \log |\hat{c}_k| + \sum_{k=1}^K \log |\hat{c}_\ell|$ for $k=1,\ldots,K$. Define the sum of the elements of $\Cm$ as $\beta \overset{\Delta}{=} \sum_{k}\sum_{\ell} \Cm_{k,\ell} = 2(K-1) \sum_{k=1}^K \log |\hat{c}_\ell|$. When $K>2$, using these equations we can obtain $|\hat{c}_k|,~k\in [K]$ as
\begin{equation}
\scalebox{0.95}{$\log |\hat{c}_k| = \frac{1}{(K-2)} \left( \sum_{\ell=1}^{K} \Cm_{k,\ell} - \frac{\beta}{2(K-1)} \right),$}
\end{equation}
which gives us an estimate of the amplitude of the CIR coefficients.\footnote{The solution for $K=1,2$ can be trivially computed and is not discussed here due to a lack of space.} Let $\theta_k$ denote the phase of $\hat{c}_k$, i.e., $\hat{c}_k = |\hat{c}_k|e^{j\theta_k}$. We estimate the CIR coefficients by treating $\hat{c}_0$ as real-valued which is of no consequences due to the fundamental phase ambiguity. This vector of coefficients, denoted by $\hat{\cv}=[|\hat{c}_0|,|\hat{c}_1|e^{j\theta_1},\ldots,|\hat{c}_{K-1}|e^{j\theta_{K-1}}]^\transp,$
is estimated as follows. For any $k=1,\ldots,K-1$, we simply find $\hat{\xi}_s$ such that $\hat{\xi}_s=\hat{\tau}_k-\hat{\tau}_0$. The corresponding estimated autocorrelation coefficient for $\hat{\xi}_i$ is given by $\hat{r}_s\approx \hat{c}_k \hat{c}_0^\ast$. By dividing $\hat{r}_s$ to $|\hat{c}_0|$ we obtain $|\hat{c}_k| e^{j\theta_k}$. 
\subsection{Resolving Ambiguities}
At this point, we have estimated the CIR up to the ambiguity arising from the shift and reflection of the support elements. To resolve this ambiguity, we rely on an observation made in \cite{vasisht2016decimeter}. As discussed in Section \ref{sec:interp}, the zero subcarrier in band $m\in [M]$ is free from the PDP phase error and contains only the constant phase error term $\psi_m$. This constant phase stems from the PLL phase offset and has the same absolute value but opposite signs on the transmitter and the receiver (see \cite{vasisht2016decimeter} Eqs. (11) and (12)), i.e.,
\begin{equation}
\begin{aligned}
\yv_{0,tx}^{(m)} = & \tilde{\hv}^{(m)}_0 e^{j\psi_m}+\zv_{0,tx}^{(m)}\\
\yv_{0,rx}^{(m)} = &\tilde{\hv}^{(m)}_0 e^{-j\psi_m}+\zv_{0,rx}^{(m)}
\end{aligned}
\end{equation}
where $\yv_{0,tx}^{(m)}$ and $\yv_{0,rx}^{(m)}$ denote CSI on the zero-subcarrier of band $m$ at the transmitter and the receiver and $\zv_{0,tx}^{(m)}$ and $\zv_{0,rx}^{(m)}$ denote their corresponding noise terms, respectively. The CFR sample $ \tilde{\hv}^{(m)}_0$ is the same on both ends due to channel reciprocity \cite{guillaud2005practical}. During frequency hopping the transmitter and the receiver send packets to each other and they can exchange their CSI measurements. We use this process to exchange interpolated zero subcarrier CSI and by multiplying them at either the receiver or transmitter WiFi device, we get
\begin{equation}\label{eq:val_exchange}
\yv'_m := \yv_{0,tx}^{(m)} \yv_{0,rx}^{(m)} = (\tilde{\hv}^{(m)}_0)^2 + \zv_m',
\end{equation} 
\vspace{-5mm}

\noindent where $\zv_m'$ denotes the noise and cross terms. The vector $\yv' = [\yv_0',\ldots,\yv'_{M-1}]^\transp\in \bC^{M}$ gives us the extra information with which we resolve the ambiguity about the shift and reflection in the support set $\hat{\Tc}$. This is done by constructing a simple hypothesis test. Let the elements of $\hat{\Tc}$ be ordered as $\hat{\tau}_0 (=0)< \ldots<\hat{\tau}_{K-1}$. In addition, we know that in an indoor environment the first delay component is within a certain limits. Let $\bar{\tau}>0$ denote this limit. Our first hypothesis is that the estimated CIR is only a shifted version of the true CIR. Let the tentative CIR representing this hypothesis $H_1$ be denoted by $f_+(\tau;\tauep) = \sum_{k=0}^{K-1} \hat{c}_k \delta (\tau-\hat{\tau}_k-\tauep),$ where by construction we made the first estimated delay component to appear at zero, i.e. $\hat{\tau}_0 = 0$ (see Algorithm \ref{Alg:supp_rec}, line 7). Also define the conjugate-reflected and shifted version of $f_+(\tau;\tauep)$ to be $f_-(\tau;\tauep) = \sum_{k=0}^{K-1} \hat{c}_k^\ast \delta (\tau+\hat{\tau}_k-\hat{\tau}_{K-1}-\tauep),$ which represents the CIR of the second hypothesis $H_2$. Define the Fourier transform of the shifted CIRs corresponding to $H_1$ and $H_2$ by a value $\tauep$ as $ F_+ (f;\tauep)  = \Fc \{ f_+(\tau ; \tauep ) \}|_f $ and $ F_- (f;\tauep)  = \Fc \{ f_-(\tau ;\tauep ) \}|_f $, respectively. The squared samples of these functions represent the squared CFR of the tentative solutions (up to a phase constant) and we will realize which hypothesis is true and which shift value $\tauep$ gives the true CIR by comparing these samples with the ones in $\yv'$. Let $\pv_+(\tauep) = [F_+^2 (f_{0,0};\tauep),\ldots,F_+^2 (f_{M-1,0};\tauep) ]^\transp$ and $\pv_-(\tauep) =[F_-^2 (f_{0,0};\tauep),\ldots,F_-^2 (f_{M-1,0};\tauep) ]^\transp$ denote the squared frequency samples of the tentative solutions for a shift $\tauep$. For the first hypothesis $H_1$, i.e. the CIR being only shifted, form the cost function $g(\tauep | H_1) = \Vert \pv_+(\tauep)  - \yv' \Vert_2$. In a similar way, for the second hypothesis $H_2$, i.e. the CIR being conjugate-reflected define the cost function as $g(\tauep |H_2) = \Vert \pv_-(\tauep)  - \yv' \Vert_2$. We select $H_1$ over $H_2$ if
\[ \underset{\tauep \in [0,\bar{\tau}]}{\min} g(\tauep |H_1) < \underset{\tauep \in [0,\bar{\tau}]}{\min} g(\tauep |H_2), \]
and $H_2$ over $H_1$ otherwise. In addition, the optimal value of the shift parameter $\tauep$ is given by $\tauep^\star = \underset{\tauep}{\arg \min }~g(\tauep |H_i),$ where $H_i$ is the winning hypothesis. After obtaining $\tauep^\star$, the eventual set of estimated delays is given by $\hat{\Tc}+\tauep^\star$. This completes the CIR estimation process and $\tauep^\star$ denotes the estimated ToF.
\begin{figure}[t]
	\centering
	\includegraphics[ width=0.4\textwidth]{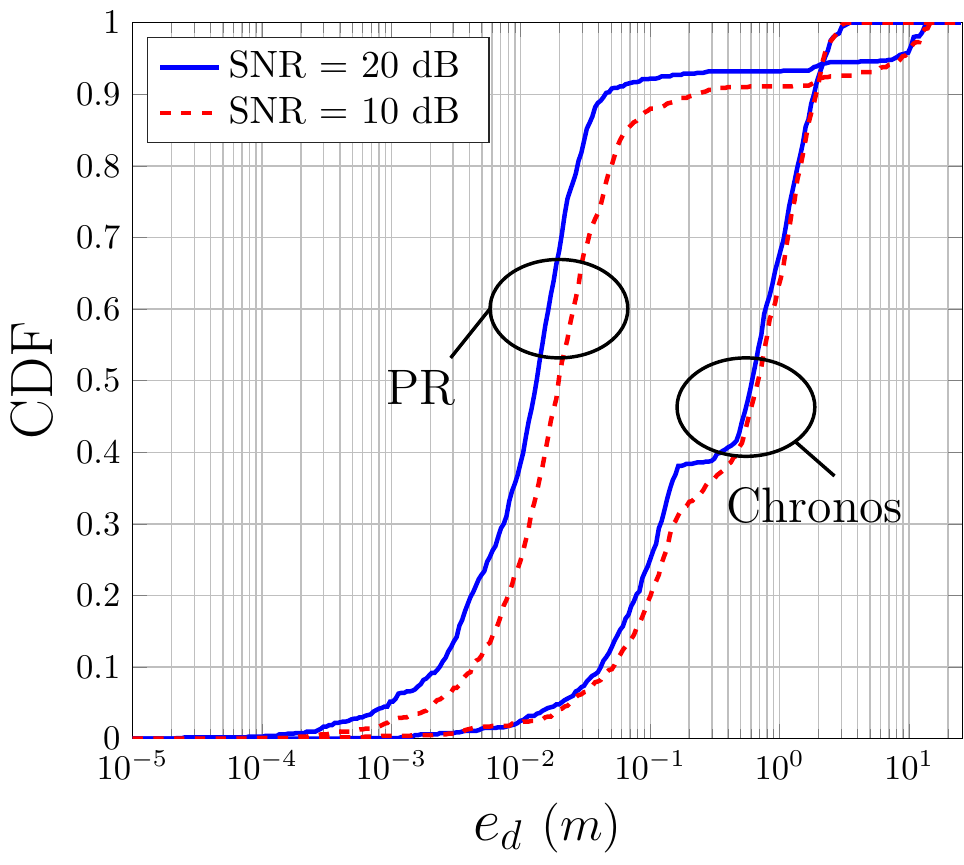}
	\caption{CDF of the ranging error for Chronos and our proposed method based on PR for SNR values of $10$ dB and $20$ dB. }
	\label{fig:CDF}
\end{figure} 
\section{Simulation Results}\label{sec:sim}
We compare our proposed method to the Chronos system developed in \cite{vasisht2016decimeter}. The method used in this system first interpolates the CSI on the zero subcarrier in each WiFi band using a cubic spline interpolator. Then, the zero-subcarrier CSI is exchanged between the WiFi device and the client to get the vector $\yv'$ in \eqref{eq:val_exchange}. This vector represents the frequency samples of the convolution of the CIR with itself. A sparse recovery method is applied to $\yv'$ to get a discretized estimate of the convolution. The first non-zero component in the estimated vector is located at twice the ToF and this gives a way to estimate the ToF and the eventual localization. 

To run the simulations we consider a CIR with $K=3$ paths whose delays are randomly located over the delay domain, such that no elements in the generated delay difference set, corresponding to different delay values, are the same. 
To each path a complex Gaussian coefficient is assigned. The coefficient assigned to a path with a larger delay has a smaller variance, since paths with larger delay usually have smaller power. We consider $M=32$ adjacent WiFi bands, each with $N=33$ subcarriers, with a carrier spacing of $f_s=312.5$ kHz. We set the parameters in programs \eqref{eq:l1_denoising_band} and \eqref{eq:sp_R_est} as $L=3N$ and $G=3MN$, respectively. Appropriate regularization scalars $\rho$ and $\lambda$ are computed as a function of the noise power using a training set prior to the experiments. Given an estimate for the first delay component for either of the methods as $\hat{\tau}_0$, the ranging error is computed as $e_d = |\tau_0 - \hat{\tau}_0|C$ (in meters), where $C$ denotes the speed of light. We run the experiment for $1000$ Monte-Carlo trials. The ranging error CDF is illustrated in Fig. \ref{fig:CDF} for two SNR values of $10$ and $20$ dB. As we can see, our proposed method outperforms Chronos in about $94\%$ of the trials, and has a ranging error less than $10$ cm in about $90\%$ of the trials, indicating a cm-level localization accuracy. Notice that even when SNR=10 dB, the PR-based method performs better than Chronos fed with CSI measurements with SNR=20 dB in about $92\%$ of the times, showing the noise resilience of our proposed method. In a small fraction of the experiments our method has a larger ranging error and performs worse compared to Chronos. These outlying results occur due to the errors in the identification of the difference set of the delays.
\vspace{-2mm}

\section{Conclusion}\label{sec:conclusion}
We proposed a novel method for indoor localization via channel impulse response estimation using WiFi multi-band CSI measurements. Our method was based on the phase retrieval of CSI magnitudes, overcoming the inherent signal phase distortions caused by the WiFi hardware. We empirically showed that this method improves the channel delay profile resolution, which in turns results in a more accurate localization of the clients in the environment in both high and low-SNR regimes.

\balance
{\small
	\bibliographystyle{IEEEtran}
	\bibliography{references}
}

\end{document}